\documentclass[twocolumn,showpacs,amsmath,amssymb]{revtex4}

\usepackage{graphicx}% Include figure files
\usepackage{dcolumn}% Align table columns on decimal point
\usepackage{bm}% bold math

\usepackage{xcolor}
\begin{document}
%%%%%%%%%%%%%%%%%%%%%%%%%%%%%%%%%%%%%%%%%%%%%%%%%%%%%%%%%%%%%%%%%%%%%%%%%%%%%%%
\title{
Detection of Macroscopic Entanglement by Correlation of 
Local Observables
}
%%%%%%%%%%%%%%%%%%%%%%%%%%%%%%%%%%%%%%%%%%%%%%%%%%%%%%%%%%%%%%%%%%%%%%%%%%%%%%%

\author{Akira Shimizu}
\email{shmz@ASone.c.u-tokyo.ac.jp}
\affiliation{
Department of Basic Science, University of Tokyo, 
3-8-1 Komaba, Tokyo 153-8902, Japan
}
\affiliation{
PRESTO, Japan Science and Technology Corporation,
4-1-8 Honcho, Kawaguchi, Saitama, Japan
}
\author{Tomoyuki Morimae}
% \email{morimae@ASone.c.u-tokyo.ac.jp}
\affiliation{
Department of Basic Science, University of Tokyo, 
3-8-1 Komaba, Tokyo 153-8902, Japan
}
\affiliation{
PRESTO, Japan Science and Technology Corporation,
4-1-8 Honcho, Kawaguchi, Saitama, Japan
}
%%%%%%%%%%%%%%%%%%%%%%%%%%%%%%%%%%%%%%%%%%%%%%%%%%%%%%%%%%%%%%%%%%%%%%%%%%%%%%%
\date{\today}
%%%%%%%%%%%%%%%%%%%%%%%%%%%%%%%%%%%%%%%%%%%%%%%%%%%%%%%%%%%%%%%%%%%%%%%%%%%%%%%
\begin{abstract}
We propose a correlation of local observables on many sites in macroscopic 
quantum systems. By measuring the correlation one can detect, if any, 
superposition of macroscopically distinct states, which we call macroscopic 
entanglement, in arbitrary quantum states that are (effectively) homogeneous.
Using this property, we also propose an index of macroscopic entanglement.
\end{abstract}
%%%%%%%%%%%%%%%%%%%%%%%%%%%%%%%%%%%%%%%%%%%%%%%%%%%%%%%%%%%%%%%%%%%%%%%%%%%%%%%
\maketitle

% Intro.

Superposition of macroscopically distinct states
has been attracting much attention
since the birth of quantum theory 
\cite{Schr,Leggett1,mermin,mermin2,SM02}.
We say a quantum state, represented by a density operator $\hat \rho$, 
is {\em entangled macroscopically}
if $\hat \rho$ has such superposition. % \cite{SM02,MSS05}.
However, the term 
`superposition of macroscopically distinct states' 
is quite ambiguous in general. % \cite{SM02,MSS05}.
For example, 
do the following states of 
a system composed of $N$ ($\gg 1$) spins 
have such superposition?
(i) 
$
| \psi_1 \rangle \equiv
\sqrt{1- 1/N} \
|\downarrow\downarrow \cdots \downarrow \rangle
+
\sqrt{1/N} \
|\uparrow\uparrow \cdots \uparrow \rangle
$,
(ii) 
$
| \psi_2 \rangle \equiv
(|\downarrow\downarrow\downarrow \cdots \downarrow \rangle
+
|\uparrow\downarrow\downarrow \cdots \downarrow \rangle
+
|\uparrow\uparrow\downarrow \cdots \downarrow \rangle
+
\cdots
+
|\uparrow\uparrow\uparrow \cdots \uparrow \rangle
)/\sqrt{N+1}
$,
and (iii) classical mixtures of macroscopically entangled states.

For {\em pure} states 
% an essentially unique 
a reasonable criterion has been given 
in Refs.~\cite{SM02,MSS05}, using which we can show that 
$| \psi_2 \rangle$ is macroscopically entangled
whereas $| \psi_1 \rangle$ is not.
Importantly, 
% it was shown that 
macroscopic entanglement 
{\em as defined by this criterion} is 
closely related to 
fundamental stabilities of quantum states
\cite{SM02}.
It was also shown that 
in quantum computers 
macroscopically entangled states are {\em always} used
to solve hard problems quickly \cite{US04,US05}.
In experiments, however, it would be hard to 
generate and confirm pure states for macroscopic systems, hence
the criterion for pure states may be difficult to apply.
Thus the following questions arise:
How can we {\em detect} macroscopic entanglement of 
an {\em unknown} state?
How can we define macroscopic entanglement for 
{\em mixed} states?

The purpose of this paper is to answer these questions.
We first show that 
macroscopic entanglement of unknown states can {\em not} 
be detected if one looks only at 
% the expectation values of 
% additive variables.
% Neither can it be detected through 
the expectation values of
low-order polynomials \cite{low-order} of additive variables
(which are fundamental macroscopic variables; see below).
Hence, it should be detected by some many-point correlations of 
local observables.
Among such correlations, we point out that
Mermin's correlation \cite{mermin,mermin2} can detect 
macroscopic entanglement only for special states.
We thus 
propose a new correlation $\hat C_{\hat A \hat \eta}$,
which 
is a function of two operators $\hat A$ and $\hat \eta$ (see below),
for general macroscopic systems composed of $N$ ($\gg 1$) sites. 
% \cite{macro}.
It can be measured 
by measuring local observables
%, two or more observables for each site,
of all sites and collecting the data thereby obtained.
For a state
represented by a density operator $\hat \rho$, 
we focus on the maximum value of
the expectation value
$\langle C \rangle = {\rm Tr} (\hat \rho \hat C_{\hat A \hat \eta})$
over all possible choices of $\hat A$ and $\hat \eta$,
and define an index $q$ of $\hat \rho$ by
\begin{equation}
\max_{\hat A, \hat \eta} 
\left( \langle C \rangle, N \right) = O(N^q).
\label{eq:q}\end{equation}
Here and after, we say that $f(N)=O(g(N))$ if
$\lim_{N\to\infty}f(N)/g(N)={\rm constant}\neq 0$.
% \cite{O(N)}.
We will show that $1 \leq q \leq 2$, and that 
it is reasonable to call states with 
$q=2$ macroscopically entangled states.
Hence, one can detect macroscopic entanglement by measuring 
$\langle C \rangle$.

{\it Basic idea --- }
%To discuss $N$ dependences, 
We consider quantum states which are
homogeneous, or effectively homogeneous as in Refs.~\cite{US04,SS03}.
We say a quantum state (or system) is {\em macroscopic} if
for every quantity of interest
the term that is leading order in $N$ gives the dominant contribution.
In general, macroscopic states are characterized by macroscopic variables, 
among which additive variables are fundamental 
because macroscopic states can be fully specified
by (a proper set of) additive variables \cite{Callen,MSS05}.
Hence, two states are {\em macroscopically distinct} iff
there is an additive variable $A$ such that its difference
is $O(N)$ between the two states.
In quantum systems, 
additive variables are represented by 
additive observables;
$ %\begin{equation}
\hat A = \sum_{l=1}^N \hat a(l),
$ %\end{equation}
where $\hat a(l)$ is a local operator at site $l$. 
Throughout this paper,
we assume that all $\hat a(l)$'s are Hermitian.
For a spin system, for example, 
such observables include
the magnetization 
$\hat M_\alpha = \sum_l \hat \sigma_\alpha(l)$
($\alpha = x, y, z$)
and
the staggered magnetization
$\hat M_\alpha^{\rm st} = \sum_l (-1)^l \hat \sigma_\alpha(l)$, 
in which $\hat a(l)=(-1)^l \hat \sigma_\alpha(l)$.
% and so on.
Note that $\hat a(l')$ for $l' \neq l$ is not necessarily 
the spatial translation of $\hat a(l)$.
To avoid mathematical complexities, we henceforth assume 
that % $\hat a(l)$ is bounded, i.e., 
$\| \hat a(l) \|$ is finite and independent of $N$, and thus
$\| \hat A \| = O(N)$.

Let $\hat A$ be an additive observable, and 
$| A \nu \rangle$ its eigenstate;
$ %\begin{equation}
\hat A | A \nu \rangle = A | A \nu \rangle,
$ %\end{equation}
where $\nu$ labels degenerate eigenstates.
According to the above argument, a quantum state 
$\hat \rho$ has more 
superposition of macroscopically distinct states, 
i.e., is more entangled macroscopically, if
$|\langle A \nu | \hat \rho | A' \nu' \rangle|$'s with 
$|A-A'| =O(N)$ are larger for a certain 
additive observable $\hat A$.
Our task is thus to propose 
% a reasonable definition of `significant' and 
a way of detecting
such $\langle A \nu | \hat \rho | A' \nu' \rangle$'s for
general $\hat \rho$.

{\it 
% Impossibility of detecting macroscopic entanglement by 
Expectation values of low-order polynomials of 
additive observables --- }
% by measurement of macroscopic observables --- }
One might expect that $\langle A \nu | \hat \rho | A' \nu' \rangle$
could be detected, if it exists, 
through the expectation value
of another additive observable $\hat B$.
Unfortunately, this is impossible for $|A-A'| =O(N)$.
For example, suppose that 
$\hat \rho = | \psi \rangle \langle \psi |$
and, neglecting degeneracies of $| A \nu \rangle$'s for simplicity, 
$| \psi \rangle = (| A_1 \rangle + | A_2 \rangle)/\sqrt{2}$,
where $|A_1 - A_2| = O(N)$.
Then, 
% $
% \langle A_1 | \hat \rho | A_2 \rangle = 1/2
% $
% for $|A_1 - A_2| = O(N)$.
% In this case, 
for any additive observable $\hat B = \sum_l \hat b(l)$,
we have 
$
{\rm Tr} (\hat \rho \hat B) 
% = 
% {1 \over 2} \langle A_1 | \hat B | A_1 \rangle
% +{1 \over 2} \langle A_2 | \hat B | A_2 \rangle
% +{1 \over 2} \left( \langle A_1 | \hat B | A_2 \rangle 
% + {\rm c.c.} \right)
= {\rm Tr} (\hat \rho_{\rm mix} \hat B)
$, 
where
$
\hat \rho_{\rm mix} 
= 
{1 \over 2}| A_1 \rangle \langle A_1 |
+{1 \over 2}| A_2 \rangle \langle A_2 |
$,
because $\hat B$ is the sum of single-site operators
and thus 
$\langle A_1 | \hat B | A_2 \rangle = 0$.

More generally, 
we recall that genuine quantum natures,
such as the violation of Bell-type inequalities, 
come from non-commutativity of observables.
For additive observables 
$\hat A = \sum_l \hat a(l)$ and $\hat B = \sum_l \hat b(l)$,
however, we have
$
\left\| [\hat A/N, \hat B/N] \right\|
=
\left\| \sum_l [\hat a(l), \hat b(l)] \right\|/N^2
\leq O(1/N).
% \to 0
$
% as $N \to \infty$.
This implies that higher accuracy of experiments is required
for larger $N$ to detect genuine quantum natures of a macroscopic 
state $\hat \rho$ 
through expectation values of $\hat A$, $\hat B$ and
$\hat A \hat B$ (and low-order polynomials \cite{low-order} of them).
In other words, % This implies that 
{\em any macroscopic states can be well described
by local classical theories if one looks only at 
such expectation values}
with non-vanishing errors
\cite{GL}.
This seems to be a foundation of macroscopic physics, 
such as thermodynamics and fluid dynamics, 
which are local classical theories.

As a simple example, let us consider 
the Clauser-Horne-Shimony-Holt (CHSH) correlation \cite{CHSH} 
of macroscopic variables.
Suppose that the system is hypothetically decomposed into two subsystems, 
each having $N/2$ sites.
Let $\hat A, \hat A'$ and $\hat B, \hat B'$ 
are additive observables of one subsystem and 
the other, respectively.
If we normalize them in such a way that their norms are $N/2$,
we may define their CHSH correlation by
$
\hat C^{\rm macro}_{\rm CHSH} \equiv
(\hat A \hat B + \hat A' \hat B - \hat A \hat B' + \hat A' \hat B')/(N/2)^2
$.
The expectation value 
$\langle C^{\rm macro}_{\rm CHSH} \rangle_{\rm cl}$ of the 
corresponding classical correlation 
satisfies the CHSH inequality 
$|\langle C^{\rm macro}_{\rm CHSH} \rangle_{\rm cl}| \leq 2$
for any local classical theories.
Since 
$\hat A/N, \hat A'/N, \hat B/N$, and $\hat B'/N$
all commute with each other in the $N \to \infty$ limit, 
we find that 
$\max_{\hat \rho, \hat A, \hat A', \hat B, \hat B'}
{\rm Tr} (\hat \rho \hat C^{\rm macro}_{\rm CHSH})
\to 2$
as $N \to \infty$,
{\em however anomalous the quantum state is}.

{\it Limitation of Mermin's correlation %for detecting macroscopic entanglement 
 --- }
The above result suggests that 
one should look at many-point correlations of {\em local} observables
in order to detect macroscopic entanglement.
% , i.e., 
% the sum of products of local observables on many sites.
Mermin proposed one of such correlations $\hat C_{\rm M}$ and proved 
a generalized Bell inequality for it, which is 
violated by an exponentially large factor $2^{(N-1)/2}$
by a `cat state,' i.e.,  
% Here, a cat state is
superposition with equal weights of {\em two} states
which are macroscopically distinct \cite{mermin}.
Since such a state is entangled macroscopically,
one might expect that 
$\langle C_{\rm M} \rangle = {\rm Tr} (\hat \rho \hat C_{\rm M})$
could be a good measure of macroscopic entanglement
if operators in $\hat C_{\rm M}$ are properly taken 
for each state \cite{mermin2}.
However, this is not the case in general.
For example, the state $| \psi_1 \rangle$ in the introduction
also violates Mermin's inequality by an exponentially large factor 
$\simeq 2^{(N -\log_2 N +1)/2}$.
However, 
this state is {\it not} entangled macroscopically 
because $q < 2$ (and $p=1$, see below).
% because $q=1$ (and $p=1$, see below).
Hence, $\langle C_{\rm M} \rangle$ can {\em not} detect 
macroscopic entanglement correctly, 
except for special states such as cat states.
We must therefore seek a new correlation.

{\it New correlation for detecting macroscopic entanglement and 
index $q$
 --- }
Let ${\cal H}$ be the Hilbert space by which a given macroscopic system 
composed of $N$ ($\gg 1$) sites is described.
Take arbitrarily 
an additive observable $\hat A$ and 
a projection operator $\hat \eta$ on ${\cal H}$, 
satisfying $\hat \eta^2 = \hat \eta$.
Using them, we define the following Hermitian operator;
\begin{equation}
\hat C_{\hat A \hat \eta} \equiv 
[ \hat A, [ \hat A, \hat \eta ] ]
=
\hat A^2 \hat \eta - 2 \hat A \hat \eta \hat A + \hat \eta \hat A^2.
\label{eq:C}\end{equation}
To see its physical meaning, 
we decompose $\hat \eta$ as
$
\hat \eta \equiv 
\sum_{j =1}^M | \phi_j \rangle \langle \phi_j |
$,
where $| \phi_j \rangle$'s are orthonormalized vectors
and $1 \leq M \leq \dim {\cal H}$.
Using eigenstates of $\hat A$, 
we obtain the expectation value 
$\langle C \rangle = {\rm Tr} \left( \hat \rho \hat C_{\hat A \hat \eta} \right)$ 
for a state $\hat \rho$ as
\begin{equation}
\langle C \rangle
=
% {1 \over 2N} 
\sum_{j =1}^M \sum_{A \nu A' \nu'}
(A-A')^2 
u^{j *}_{A \nu} \langle A \nu |\hat \rho | A' \nu' \rangle u^{j}_{A' \nu'},
\label{eq:<C>-Anu}
\end{equation}
where $u^j_{A \nu} \equiv \langle A \nu | \phi_j \rangle$.
For a given state $\hat \rho$,  
we focus on  the $N$ dependence of 
the maximum value 
$\max_{\hat A, \hat \eta} \langle C \rangle$ 
for all possible choices of $\hat A$ and $\hat \eta$,
and define an index $q$ by Eq.~(\ref{eq:q}).
By definition, $q \geq 1$.
As we will show shortly, 
the equality is satisfied, e.g., by
% minimum value $q=1$ is taken, e.g., by 
every separable state
(i.e., classical mixture of product states).
On the other hand, 
we find that 
$q \leq 2$ % $\max_{\hat A, \hat \eta} \langle C \rangle \leq O(N^2)$.
because
$ %\begin{equation}
% \left| {\rm Tr} \left( \hat \rho \hat C_{\hat A \hat \eta} \right) \right|
\left| \langle C \rangle \right|
\leq % {1 \over 2N} 
\| [ \hat A, [ \hat A, \hat \eta ] ] \|
%\left\| [ \hat A, [ \hat A, \hat \eta ] ] \right\|
\leq 4 %{2 \over N} 
\| \hat A \|^2 \left\| \hat \eta \right \| 
= O(N^2), 
$ %\end{equation}
where we have used $\| \hat A \| = O(N)$ and
$\| \hat \eta \| = 1$.
It is seen from Eq.~(\ref{eq:<C>-Anu}) that 
$\hat \rho$ has a larger value of 
$\max_{\hat A, \hat \eta} \langle C \rangle$ 
when $|\langle A \nu |\hat \rho | A' \nu' \rangle|$'s with  
$|A-A'| = O(N)$ are larger.
Since such matrix elements represents 
quantum coherence between macroscopically distinct states, 
it is reasonable to call $\hat \rho$ with the maximum value $q=2$ 
a {\em macroscopically entangled state}.
Note that 
the minimum value $q=1$ is taken also by the random state 
$\hat \rho = \hat 1/\dim {\cal H}$,
for which $\langle C \rangle = 0$.
Hence, the index $q$ of macroscopic entanglement
classifies separable states, 
for which quantum coherence exists only within each site,
and the random state,
for which any quantum coherence is absent, 
as a single group.
This is reasonable because 
they do not have macroscopic entanglement at all.

To sum up, 
the index $q$ of macroscopic entanglement, defined by Eq.~(\ref{eq:q}),
ranges over $1 \leq q \leq 2$.
We say $\hat \rho$ is macroscopically entangled if $q=2$, whereas
states with $q<2$ may be entangled but not macroscopically,
among which states with $q=1$ are 
similar to separable states
%the random state $\hat 1/\dim {\cal H}$
%% classified into the same group 
in view of macroscopic entanglement.

{\it Properties of $q$ for pure states --- }
For pure states, a reasonable %the essentially unique 
index $p$ of 
macroscopic entanglement was given in Refs.~\cite{SM02,MSS05} as
$\max_{\hat A} \langle \psi | (\Delta \hat A)^2 | \psi \rangle = O(N^p)$,
where 
$\Delta \hat A \equiv \hat A - \langle \psi | \hat A | \psi \rangle$
and % $p$ ranges over 
$1 \leq p \leq 2$.
We now investigate the relation between $q$ and $p$ for pure states.

If $\hat \rho$ is a pure state $| \psi \rangle \langle \psi |$,
we can easily show that 
$\hat \eta | \psi \rangle \neq 0$ is necessary to maximize
$\langle C \rangle$.
Furthermore, any $\hat \eta$ such that $\hat \eta | \psi \rangle \neq 0$
can be expressed as
$ %\begin{equation}
\hat \eta = | \phi \rangle \langle \phi | 
+
\sum_{j =2}^M | \phi'_j \rangle \langle \phi'_j |,
$ %\end{equation}
where 
$| \phi \rangle \equiv 
\hat \eta | \psi \rangle / \| \hat \eta | \psi \rangle \|$,
$\langle \psi |\phi'_j \rangle = \langle \phi |\phi'_j \rangle = 0$ and
$\langle \phi'_j | \phi'_{j'} \rangle = \delta_{j,j'}$.
Using this expression, we have
\begin{eqnarray}
\langle C \rangle
&=& % {1 \over 2N} 
\left(
\langle \phi | \hat A^2 | \psi \rangle \langle \psi | \phi \rangle 
+ {\rm c.c.} \right)
\nonumber\\
& & 
- 2 %{1 \over N} 
\left| \langle \phi | \hat A | \psi \rangle \right|^2
- 2 %{1 \over 2N} 
\sum_{j=2}^M 
\left| \langle \phi'_j | \hat A | \psi \rangle \right|^2.
\end{eqnarray}
Since this becomes maximum when $M=1$, we find
% Hence,
\begin{equation}
\max_{\hat \eta} 
% \max_{\hat A, \hat \eta} 
\langle C \rangle
=
% \max_{\hat A, | \phi \rangle} 
\max_{| \phi \rangle} 
\langle \phi |
[\hat A, [\hat A, | \psi \rangle \langle \psi |]] 
| \phi \rangle.
\label{eq:maxCpure}\end{equation}
Therefore, 
$ %\begin{equation}
\max_{\hat A, \hat \eta} 
\langle C \rangle
\geq 
\max_{\hat A} 
\langle \psi |
[\hat A, [\hat A, | \psi \rangle \langle \psi |]] 
| \psi \rangle
=
2
\max_{\hat A} \langle \psi | (\Delta \hat A)^2 | \psi \rangle,
$ %\end{equation}
from which we immediately find that
{\em if $p=2$ then $q=2$}, and {\em if $q=1$ then $p=1$}.
We also note that Eq.~(\ref{eq:maxCpure}) 
implies that 
$\max_{\hat \eta} \langle C \rangle$
is the maximum eigenvalue of the Hermitian operator
$[\hat A, [\hat A, | \psi \rangle \langle \psi |]]$.
If we denote an eigenvector corresponding to 
the maximum eigenvalue by $| \phi_A \rangle$,
we have 
\begin{eqnarray}
&& \hspace{-7mm}
% 2N 
\max_{\hat A, \hat \eta} \langle C \rangle
=
\max_{\hat A} \langle \phi_A |
[\hat A, [\hat A, | \psi \rangle \langle \psi |]] 
| \phi_A \rangle
\nonumber\\
&&  \hspace{-5mm} = 
\max_{\hat A} \mbox{Tr}\left(
[|\phi_A\rangle\langle\phi_A|,A]^\dagger
[|\psi\rangle\langle\psi|,A]\right)
\nonumber\\
&& \hspace{-5mm} \le 
2 \left[ \max_{\hat A} 
\langle\phi_A| (\Delta_{\phi_A} \hat A)^2 | \phi_A \rangle
\right]^{1 \over 2}
\hspace{-1mm}
\left[ \max_{\hat A'} 
\langle\psi| (\Delta_\psi \hat A')^2 | \psi \rangle
\right]^{1 \over 2} \hspace{-3mm},
\label{eq:leqPP}\end{eqnarray}
where 
we have used the Cauchy-Schwartz inequality,
$
|\mbox{Tr} (\hat J^\dagger \hat K)|
\leq 
[\mbox{Tr} (\hat J^\dagger \hat J)]^{1/2}
[\mbox{Tr} (\hat K^\dagger \hat K)]^{1/2}
$,
and 
$\Delta_{\phi_A} \hat A \equiv \hat A - \langle \phi_A | \hat A | \phi_A \rangle$,
$\Delta_\psi \hat A' \equiv \hat A' - \langle \psi | \hat A' | \psi \rangle$.
We thus find that 
{\em if $q=2$ then $p=2$}
and that {\em if $p=1$ then $q \leq 1.5$}.
Moreover, 
since 
$| \phi_A \rangle$
is an eigenvector of
$[\hat A, [\hat A, | \psi \rangle \langle \psi |]]$,
% and $\hat A = \sum_l \hat a(l)$,
it is given by a linear combination 
$| \phi_A \rangle 
=
x | \psi \rangle 
+ \sum_l y_l \hat a(l) | \psi \rangle 
+ \sum_{l,l'} z_{ll'} \hat a(l) \hat a(l') | \psi \rangle
$.
This implies that $| \phi_A \rangle$ is obtained from 
$| \psi \rangle$ by adding one- and two-particle excitations.
For any product state
$| \psi \rangle = \bigotimes_{l=1}^N | \psi_l \rangle$, 
addition of such microscopic excitations 
does not change the value of the index $p$ 
of macroscopic entanglement \cite{SM02,MSS05}.
Thus, from inequality (\ref{eq:leqPP}), 
we find that 
{\em $q=1$ for any product state}.
% 
% Since addition of such microscopic excitations 
% does not change the value of the index $p$ 
% of macroscopic entanglement \cite{SM02,MSS05}, 
% $p=1$ for $| \phi \rangle$ if $p=1$ for $| \psi \rangle$.
% Thus, from inequality (\ref{eq:leqPP}), 
% we find that {\em if $p=1$ then $q=1$}.
% In particular, 
% {\em $q=1$ for any product state} 
% $| \psi \rangle = \bigotimes_{l=1}^N | \psi_l \rangle$
% because $p=1$.
% 

To sum up, 
we have found that 
$q=1 \Rightarrow p=1$, 
$p=1 \Rightarrow q \leq 1.5$
% $p=1 \Leftrightarrow q=1$
and that $p=2 \Leftrightarrow q=2$,
for pure states.

{\it Properties of $q$ for mixed states --- }
The above results demonstrate that  
$q$ is a natural generalization of $p$, which was defined only 
for pure states \cite{SM02,MSS05}.
We now present basic properties of $q$ for mixed states.

{\em Any mixture 
$\hat \rho 
= \sum_\lambda \rho_\lambda | \psi_\lambda \rangle \langle \psi_\lambda |
$
of pure states $| \psi_\lambda \rangle$'s with $q = 1$ has $q = 1$}.
In fact, 
$
\max_{\hat A, \hat \eta} \langle C \rangle
\leq 
\sum_\lambda \rho_\lambda \max_{\hat A, \hat \eta} 
\langle \psi_\lambda | \hat C_{\hat A, \hat \eta} | \psi_\lambda \rangle
= 
 \sum_\lambda \rho_\lambda O(N) = O(N).
$
In particular, {\em $q=1$ for separable states}
since $q=1$ for product states.
% , i.e., 
% classical mixtures of product states.
%
On the other hand, 
{\em mixtures of pure states $| \psi_\lambda \rangle$'s with $q=2$ do not 
necessarily have $q=2$}.
A simple example 
for an $N$-spin system %array of spin-${1 \over 2}$ systems
is the state with $\rho_\pm = 1/2$ and
$
| \psi_\pm \rangle
=
(
| \downarrow \rangle^{\otimes N} 
\pm
| \uparrow \rangle^{\otimes N} 
)/\sqrt{2}
$.
Then, 
$
\hat \rho_{\rm ex1} \equiv 
{1 \over 2} | \psi_+ \rangle \langle \psi_+ |
+
{1 \over 2} | \psi_- \rangle \langle \psi_- |
$ 
is equal to 
$
{1 \over 2} (| \downarrow \rangle \langle \downarrow |)^{\otimes N} 
+ {1 \over 2} (| \uparrow \rangle \langle \uparrow |)^{\otimes N}
$,
which is a classical mixture of product states,
and thus $q=1$.

It is interesting to clarify the conditions 
for $q=2$ for mixtures of states with $q=2$.
A {\em sufficient condition} is as follows.
Suppose that for an additive operator $\hat A$ we have pure states
$| \psi_1 \rangle$, $| \psi_2 \rangle$, $\cdots$
%$| \psi_{\lambda_1} \rangle$, $| \psi_{\lambda_2} \rangle$, $\cdots$
such that 
\begin{eqnarray}
&&
\langle \psi_\lambda | \psi_{\lambda'} \rangle = \delta_{\lambda, \lambda'}
\mbox{ for } \lambda, \lambda' = 1, 2, \cdots,
\label{eq:cond1}\\
&&
\langle \psi_\lambda | \hat A | \psi_{\lambda'} \rangle = 0
\mbox{ for } \lambda \neq \lambda',
\label{eq:cond2}\\
&&
\langle \psi_\lambda | (\Delta_\lambda \hat A)^2 | \psi_\lambda \rangle
= O(N^2) 
% \mbox{ hence } q=p=2 
\mbox{ for } \lambda \leq \Lambda,
%\mbox{ (hence $q=p=2$)}
\label{eq:cond3}\\
&&
\langle \psi_\lambda | (\Delta_\lambda \hat A)^2 | \psi_\lambda \rangle
< O(N^2) 
\mbox{ for } \lambda > \Lambda,
\label{eq:cond4}\end{eqnarray}
% for $\lambda = 1, 2, \cdots$,
where 
$\Delta_\lambda \hat A \equiv
\hat A - \langle \psi_\lambda | \hat A | \psi_\lambda \rangle$
and $\Lambda$ is a positive integer.
% ($\lambda = 1, 2, \cdots$).
Consider classical mixtures of these states,
$\hat \rho 
= \sum_\lambda \rho_\lambda | \psi_\lambda \rangle \langle \psi_\lambda |$,
where  $\rho_\lambda$'s are real numbers 
such that $0 \leq \rho_\lambda \leq 1$ and 
$\sum_\lambda \rho_\lambda = 1$.
If 
\begin{equation}
\lim_{N \to \infty} \sum_{\lambda \leq \Lambda} \rho_\lambda \neq 0,
% \mbox{ and independent of } N,
\label{eq:cond5}\end{equation}
then any such mixtures have $q=2$, hence are entangled macroscopically.
In fact, if we take 
$\hat \eta = \sum_\lambda | \psi_\lambda \rangle \langle \psi_\lambda |$,
we find 
$ %\begin{equation}
\langle C \rangle
=
2 
\sum_\lambda \rho_\lambda
\langle \psi_\lambda | (\Delta_\lambda \hat A)^2 | \psi_\lambda \rangle
= O(N^2),
$ %\end{equation}
hence $q=2$. % $\max \langle C \rangle = O(N^2)$.

For example, 
let 
$
| \psi_\lambda \rangle
=
{1 \over \sqrt{2}}
| \downarrow \rangle^{\otimes (\lambda-1)} | \uparrow \rangle
| \downarrow \rangle^{\otimes (N-\lambda)} 
+
{1 \over \sqrt{2}}
| \uparrow \rangle^{\otimes (\lambda-1)} | \downarrow \rangle
| \uparrow \rangle^{\otimes (N-\lambda)}
$
for $\lambda = 1, 2, \cdots, N$.
Then, conditions (\ref{eq:cond1})-(\ref{eq:cond3}) are all satisfied 
for $\hat A = \hat M_z = \sum_l \hat \sigma_z(l)$ and $\Lambda = N$.
Therefore, % According to the above result, 
any mixtures of these states, such as
$\hat \rho_{\rm ex2} \equiv 
(1/N) \sum_{\lambda=1}^N | \psi_\lambda \rangle \langle \psi_\lambda |$,
are entangled macroscopically, i.e., $q=2$.
This may be understood by noting that 
such mixtures are mixtures of 
the 'same sort' of superpositions of macroscopically distinct states
in the sense that 
all $| \psi_\lambda \rangle$'s 
are superpositions of states with $M_z = \pm(N-2)$.

A more instructive example is the case where
$
| \psi_\lambda \rangle 
\equiv (| \lambda \rangle +|\bar{\lambda}\rangle)/ {\sqrt{2}}
$,
where $|\lambda\rangle$ ($|\bar{\lambda}\rangle$) 
is an arbitrary state in which $\lambda$ spins are up (down) and 
$N-\lambda$ spins are down (up).
If we limit the range of $\lambda$ over, say, 
$1 \leq \lambda \leq N/3$, then
conditions (\ref{eq:cond1})-(\ref{eq:cond3}) are all satisfied 
for $\hat A = \hat M_z$ and $\Lambda = N/3$.
Therefore, any mixtures of these states, such as
$
\hat \rho_{\rm ex3} \equiv 
(3/N) \sum_{\lambda=1}^{N/3} 
| \psi_\lambda \rangle \langle \psi_\lambda |
$,
are entangled macroscopically, i.e., $q=2$.
Intuitively, 
such mixtures are mixtures of 
the same sort of superpositions of macroscopically distinct states
in the sense that 
all $| \psi_\lambda \rangle$'s 
are superpositions of states with positive and negative $M_z$.

Furthermore, 
$
\hat \rho'_{\rm ex2} \equiv w \hat \rho_{\rm ex2} + (1-w) \hat \rho_{\rm ex1}
$
and 
$
\hat \rho'_{\rm ex3} \equiv w \hat \rho_{\rm ex3} + (1-w) \hat \rho_{\rm ex1}
$
also have $q=2$ if $w > 0$ and independent of $N$,
because 
$
| \downarrow \rangle^{\otimes N} 
$
and
$
| \uparrow \rangle^{\otimes N} 
$
satisfy the above conditions for 
$| \psi_\lambda \rangle$'s with $\lambda > \Lambda$.

{\it Measurement of $\langle C \rangle$ by local measurements ---}
When detecting entanglement 
of two particles by measuring the CHSH correlation,
$
\hat C_{\rm CHSH} = 
\hat a(\theta) \hat b(\phi) + \hat a(\theta') \hat b(\phi) 
-\hat a(\theta) \hat b(\phi') + \hat a(\theta') \hat b(\phi')
$,
one does {\em not} measure it using 
a {\em single} experimental setup, which performs a {\em global} (non-local)
measurement.
Instead, one measures 
$\hat a$'s and $\hat b$'s locally and simultaneously,
which are observables of one particle and the other, respectively.
% \cite{Aspect}.
Since % $[\hat a(\theta), \hat a(\theta')] \neq 0$, 
$\hat a(\theta)$ and $\hat a(\theta')$ %($\hat b(\phi)$ and $\hat b(\phi')$)
cannot be measured 
simultaneously because $[\hat a(\theta), \hat a(\theta')] \neq 0$, 
% they do not commute with each other,
they should be measured independently using 
different experimental setups,
and similarly for $\hat b(\phi)$ and $\hat b(\phi')$.
That is, one performs % many runs of 
local measurements with various setups.
By collecting the data of such local measurements, 
one can obtain the expectation values of all terms in $\hat C_{\rm CHSH}$,
and hence the value of $\langle C_{\rm CHSH} \rangle$.

In a similar manner, 
one can obtain $\langle C \rangle$ 
by measuring local observables with various setups 
and collecting the data thereby obtained.
This might be obvious because in general any Hermitian operator on 
${\cal H} = \bigotimes_l {\cal H}_l$,
where ${\cal H}_l$ is the 
local Hilbert space of site $l$, 
can be expressed as the sum of products of local Hermitian 
operators. % each operating on ${\cal H}_l$.
However, we show it
in such a way that local observables to be measured can be seen easily.
Let 
% ${\cal H} = \otimes_l {\cal H}_l$, where ${\cal H}_l$ is the 
% local Hilbert space of site $l$, and 
$| a_l \mu_l \rangle \in {\cal H}_l$ be 
an eigenvector of $\hat a(l)$; 
$\hat a(l) | a_l \mu_l \rangle = a_l | a_l \mu_l \rangle$,
where $\mu_l$ labels degenerate eigenvectors.
We can take $| A \nu \rangle = \bigotimes_l | a_l \mu_l \rangle$,
where $A = \sum_l a_l$. %and $\mu = (\mu_1, \mu_2, \cdots, \mu_N)$.
Hence, denoting 
${\bm a} = (a_1, a_2, \cdots, a_N)$ and 
${\bm \mu} = (\mu_1, \mu_2, \cdots, \mu_N)$,
we can express $\hat C_{\hat A \hat \eta}$ as
\begin{eqnarray}
\hat C_{\hat A \hat \eta} &=&
\sum_{j =1}^M \sum_{{\bm a} {\bm \mu} {\bm a}' {\bm \mu}'}
\left( \sum_{l'} (a_{l'} -a'_{l'}) \right)^2 
u^{j}_{{\bm a}' {\bm \mu}'}  u^{j *}_{{\bm a} {\bm \mu}}
\nonumber\\
& & \ \ \times \bigotimes_l 
\left( \hat \varphi'_{a'_l \mu'_l a_l \mu_l}(l)
+i \hat \varphi''_{a'_l \mu'_l a_l \mu_l}(l) \right),
\label{eq:Cbylocal}\end{eqnarray}
where 
$\hat \varphi'_{a'_l \mu'_l a_l \mu_l}(l)
\equiv (| a'_l \mu'_l \rangle \langle a_l \mu_l | + {\rm h.c.})/2$
and 
$\hat \varphi''_{a'_l \mu'_l a_l \mu_l}(l)
\equiv (| a'_l \mu'_l \rangle \langle a_l \mu_l | - {\rm h.c.})/(2i)$
are local Hermitian operators on ${\cal H}_l$.
By expanding Eq.~(\ref{eq:Cbylocal}), we obtain a polynomial of 
$\hat \varphi'(l)$'s and $\hat \varphi''(l)$'s,
i.e., 
the sum of products of local observables.
Therefore, 
$\langle C \rangle$ can be measured 
by measuring such local observables of each terms
(using proper experimental setups for each)
and collecting the data thereby obtained.

The operators $\hat \varphi'(l)$'s and $\hat \varphi''(l)$'s,
which we denote $\hat {\bm \varphi}$,  
and the numbers ${\bm a}$, ${\bm \mu}$
% which are functions of $\hat A$ and $\hat \eta$,
in Eq.~(\ref{eq:Cbylocal}) 
correspond to % the set of
$\hat a$, $\hat b$, $\theta, \theta', \phi, \phi'$
of $\hat C_{\rm CHSH}$.
To find the value of $q$, one should seek a particular set of 
$\hat {\bm \varphi}$, ${\bm a}$, ${\bm \mu}$
% $\hat \varphi'$'s, $\hat \varphi''$'s, ${\bm a}, {\bm \mu}$ 
that maximizes $\langle C \rangle$ 
(or gives the same order of magnitude of $\langle C \rangle$
as the maximum value).
If the state $\hat \rho$ is unknown, 
one should perform experiments for various choices of 
$\hat {\bm \varphi}$, ${\bm a}$, ${\bm \mu}$,
and thereby find the maximum value of $\langle C \rangle$.
This situation is the same as the case of detecting 
the violation of the CHSH inequality
of two particles by an unknown state,
where one should perform experiments for various choices of 
$\hat a$, $\hat b$, $\theta, \theta', \phi, \phi'$.
In many practical experiments, however, one tries to generate some
target state with a prescribed $\hat \rho$.
In such a case, one can theoretically find 
$\hat A$ and $\hat \eta$ that should give the maximum value of 
$\langle C \rangle$ \cite{MSunpub}.
Then, one needs to measure $\langle C \rangle$ 
only for $\hat {\bm \varphi}$, ${\bm a}$, ${\bm \mu}$
corresponding to such $\hat A$ and $\hat \eta$.

{\it Conversion of states with $q<2$ to states with $q=2$ --- }
Entanglement is often defined in terms of possibility 
of converting a state in question to 
another state which is manifestly entangled \cite{NC}.
% We discuss this point briefly.
In the present case, 
it is possible to convert 
$| \psi_1 \rangle$ in the introduction, 
which has $q=1.5$, 
to a cat state,
which has $q=2$, 
by a {\em single}-spin projective measurement.
% Although this is a very interesting observation, 
However, its success probability tends to vanish with increasing $N$.
In our opinion,
it is natural to exclude such rare events to 
{\em define} macroscopic entanglement, 
and to interpret the above possibility as 
an interesting possibility with a very small but non-vanishing
(for finite $N$) success probability.
% Although this is a very interesting observation, 

{\it Possible experiments --- }
It is very interesting to detect macroscopic entanglement experimentally.
One way of producing states with $q=2$ is
to cool a symmetry-breaking system whose order parameter does not 
commute with the Hamiltonian, such as the Heisenberg 
antiferromagnet on a two-dimensional square lattice \cite{MSS05}.
If the temperature can be made lower
than the energy difference % $\Delta E$ 
between 
the exact ground state (which is symmetric \cite{SM02,MSS05,pre01})
and the symmetry-breaking vacuum,
then the equilibrium density operator becomes 
a macroscopically entangled state \cite{MSunpub}.
Another way % of producing states with $q=2$ 
may be to use quantum computers,
in which one can manipulate quantum states rather freely \cite{NC}, 
as a playground of many-body physics.

We thank M. Koashi and H. Tasaki for discussions.

%%%%%%%%%%%%%%%%%%%%%%%%%%%%%%%%%%%%%%%%%%%%%%%%%%%%%%%%%%%%%%%%%%%%%%%%%%%%%%

\begin{thebibliography}{00}
\bibitem{low-order}
Low-order polynomials mean here $m$-th order polynomials with $m=O(1)$, 
although we expect that the same can be said for any $m$ such that
$m \ll N$.
\bibitem{GL}
For example, % it is well-known that 
macroscopic properties of 
superconductors are well described by the Ginzburg-Landau theory,
which is surely a local classical theory, although 
its macroscopic variable is called 
the `macroscopic wavefunction.'
\end{thebibliography}
\end{document}